\documentclass[conference]{IEEEtran}
\IEEEoverridecommandlockouts
\usepackage{cite}
\usepackage{amsmath,amssymb,amsfonts}
\usepackage{algorithmic}
\usepackage{algorithm}
\usepackage{graphicx}
\usepackage{textcomp}
\usepackage{xcolor}
\usepackage{url}
\usepackage{subfloat}
\makeatletter
\newcommand\fs@betterruled{%
  \def\@fs@cfont{\bfseries}\let\@fs@capt\floatc@ruled
  \def\@fs@pre{\vspace*{5pt}\hrule height.8pt depth0pt \kern2pt}%
  \def\@fs@post{\kern2pt\hrule\relax}%
  \def\@fs@mid{\kern2pt\hrule\kern2pt}%
  \let\@fs@iftopcapt\iftrue}
\floatstyle{betterruled}
\restylefloat{algorithm}
\makeatother

\def\BibTeX{{\rm B\kern-.05em{\sc i\kern-.025em b}\kern-.08em
    T\kern-.1667em\lower.7ex\hbox{E}\kern-.125emX}}
\begin{document}

\title{Multi-domain Network Slice Partitioning: A Graph Neural Network Algorithm}

\author{\IEEEauthorblockN{Zhouxiang Wu\textsuperscript{1}, Genya Ishigaki\textsuperscript{2}, Riti Gour\textsuperscript{3}, Congzhou Li\textsuperscript{1}, Divya Khanure\textsuperscript{1} and Jason P. Jue\textsuperscript{1}}
    \IEEEauthorblockA{
        1. Department of Computer Science, The University of Texas at Dallas, Richardson, Texas 75080, USA\\
        2. Department of Computer Science, San Jose State University, San Jose, CA 95192, USA\\
        3. Department of Computer Electronics and Graphic Technology,\\ Central Connecticut State University, New Britain, Connecticut 06050, USA
    }
}

\maketitle

\begin{abstract}
In the context of multi-domain network slices, multiple domains need to work together to provide a service. The problem of determining which part of the service fits within which domain is referred to as slice partitioning.
The partitioning of multi-domain network slices poses a challenging problem, particularly when striving to strike the right balance between inter-domain and intra-domain costs, as well as ensuring optimal load distribution within each domain. To approach the optimal partition solution while maintaining load balance between domains, a framework has been proposed. This framework not only generates partition plans with various characteristics but also employs a Graph Neural Network solver, which significantly reduces the plan generation time. The proposed approach is promising in generating partition plans for multi-domain network slices and is expected to improve the overall performance of the network.
\end{abstract}
\section{Introduction}
Network slicing \cite{samdanis2016network} has been developed to address users' diverse requirements by leveraging virtual network functions (VNF) and software-defined networking (SDN). This approach creates isolated network environments based on existing network infrastructure. Traditional network slices have typically been confined to single or two adjacent domains, which often fail to achieve globally optimal resource management. Consequently, multi-domain end-to-end network slicing has emerged as a solution.

Multi-domain network slices \cite{taleb2019multi} are commonly managed by a global coordinator who orchestrates resources across various domains. If a domain operator shares their domain's detailed information with the global coordinator, the problem becomes akin to the conventional VNF embedding problem \cite{afolabi2018network} but on a larger scale. However, in cases where domain operators do not divulge such information, the global coordinator must employ inference techniques to map VNFs to the respective domains.

Consequently, the current mapping problem encompasses two stages. In the first stage, the global coordinator partitions the multi-domain network slice into distinct segments and assigns each segment to a domain, considering that multiple domains may be capable of supporting a given VNF. During the second stage, the domain operators develop embedding plans based on their assigned segments. The global coordinator then selects a set of embedding plans and stitches them together to deliver end-to-end services. In this paper, our primary focus lies in the first stage: efficiently partitioning network slices to minimize resource costs.

Partitioning network slices presents several challenges. The first challenge involves minimizing resource costs, which can be divided into two components: the resource cost within the domain and the resource cost between domains. Placing all elements within a single domain would undoubtedly reduce the resource cost between domains to zero; however, this would result in a high resource cost within the domain. Striking a balance between these two types of costs is essential. Another challenge concerns efficiency. As the global coordinator must perform the partitioning task for each incoming network slice request, a lengthy process time would contribute to increased queueing wait times for establishing and deploying new network slices. Additionally, maintaining load balance among domains is critical, as imbalanced arrangements can lead to resource waste. 

Our study presents a formalization of the partition problem using an Integer Programming approach. Furthermore, we propose three distinct methodologies to address this problem. The remainder of this paper is structured as follows: Section \ref{sec:related_work} provides a summary of the related work. Section \ref{sec:problem_formulation} presents the problem formulation. In Section \ref{sec:methodology}, we propose three algorithms to address the problem. Simulation experiments and result analyses are conducted in Section \ref{sec:experiments}. Lastly, we draw our conclusions in Section \ref{sec:conclusion}.
\section{Related Work}
\label{sec:related_work}
In previous work \cite{icc2023}, we designed an architecture that enabled the global coordinator to select allocation plans generated by domain operators. However, that approach employed a fixed partitioning strategy. In this paper, we transition from a fixed strategy to a more dynamic and objective-oriented partitioning plan. 
There is a dearth of literature on the subject of network slice partitioning. Therefore, in this study, we provide an overview of the relevant research pertaining to the embedding of multi-domain Virtual Network Function Forwarding Graphs (VNF-FGs) and the associated load-balancing strategies.

In \cite{8845184}, the authors note the complexities and challenges associated with multi-domain non-cooperative scenarios. They propose a deep reinforcement learning-based approach for VNF-FG embedding to address these issues. The results reveal insights into the behavior of non-cooperative domains and demonstrate the efficiency of their approach, which enables automatic inter-domain load balancing. 
The paper \cite{8494813} proposes a model of the adaptive and dynamic Virtual Network Function - Forwarding Graph (VNF-FG) allocation problem, an integer linear programming (ILP) optimization problem, and a heuristic algorithm for allocating multiple VNF-FGs. They also propose a decentralized optimization approach for cooperative multi-operator scenarios. The results confirm that the proposed algorithms optimize network utilization while limiting the number of VNF reallocations that could interrupt network services. 
The paper \cite{9833394} proposes a coordination-free algorithm that supports non-functional dependencies and offers strong eventual consistency. The algorithm has two variants, one preventive and one corrective, and the paper proves the correctness of both variants. The proposed algorithm outperforms the state of the art in terms of consistency and allowing orchestrators to reject ongoing reconfigurations without a significant impact on performance. 
\section{Problem Formulation}
\label{sec:problem_formulation}
\subsection{Multi-domain Network Infrastructure}
The aim of this study is to investigate scenarios in which multiple domains collaborate to provision an end-to-end network slice. We assume that each type of domain is both contiguous and unique. To represent these domains, we use a list with $M$ elements, where $M$ is the number of domains. For example, there might be three distinct but contiguous domains: RAN, edge, and core. Our work can be extended easily to scenarios with multiple domains of each type by integrating parallel domains. Each domain is composed of non-disclosable resources, such as computation nodes and data transmission links. Importantly, the topology information and resource allocation within each domain are not accessible to the global coordinator due to access restrictions. Instead, each domain $m \in M$ provides the global coordinator with estimates of the costs for CPU, RAM, bandwidth, and latency for its supported Virtual Network Functions (VNFs), denoted as $c_{CPU}(m)$, $c_{RAM}(m)$, $c_{Link}(m)$, and $l(m)$, respectively.
Using this information, the global coordinator provisions the network slice partition plans across multiple domains. Domains also disclose their in-bound and out-bound points to the global coordinator. Furthermore, the global coordinator has access to information on the bandwidth on links between different domains and the associated costs per unit of bandwidth on an inter-domain link between domain $m_i$ and $m_j$, which is given by $c_{Link}(m_i,m_j)$.
\subsection{VNF-FG}
We make the assumption that the network slice can be accurately represented as a VNF-FG, denoted by $G$. This VNF-FG maintains a set of nodes, denoted by $N$, and a set of edges, denoted by $E$.
Each node $n \in N$ is characterized by two numerical values: $CPU_n$ and $RAM_n$, which respectively represent the required level of CPU and RAM for the corresponding VNF. The edge $e \in E$ is characterized by the required bandwidth necessary for successful data transmission between two VNFs in the network slice. Given the complexity in determining end-to-end latency for a graph, we work under the assumption that each VNF-FG has a predefined total latency requirement $l$.
\subsection{VNF-FG Partitioning Problem Formulation}

The partitioning problem can be formulated as a node classification problem, where a binary variable $X_n^m$ indicates whether node $n \in N$ is allocated to domain $m$. Specifically, $X_n^m$ is defined as:

\begin{equation}
    X_n^m = 
    \begin{cases}
    1, & \text{if } n \text{ is allocated to } m\\
    0, & \text{otherwise}
    \end{cases}.
\end{equation}

The objective is to minimize the cost generated by the domains, as well as the cost generated by inter-domain connections. The cost generated by the domains can be formulated as a function $DC$:

\begin{equation}
    \begin{aligned}
    DC & = \sum_{n \in N} \sum_{m \in M} X_n^m \times \\
       & \{ c_{CPU}(m) \times CPU_n + c_{RAM}(m) \times RAM_n \}.
    \end{aligned}
\end{equation}

Similarly, the link cost generated by domains can be formulated as a function $DL$:

\begin{equation}
    \begin{aligned}
        DL &= \sum_{n_i \in N} \sum_{n_j \in N} \sum_{m \in M} X_{n_i}^m \times X_{n_j}^m \times bw(n_i, n_j) \times c_{Link}(m).
    \end{aligned}
\end{equation}

The inter-domain link cost can be formulated as a function $IC$:

\begin{equation}
    \begin{aligned}
        IC &= \sum_{n_i \in N} \sum_{n_j \in N} \sum_{m_i \in M} \sum_{m_j \in M \backslash \{m_i\}} \\ 
        & X_{n_i}^{m_i} \times X_{n_j}^{m_j} \times bw(n_i, n_j) \times c_{Link}(m_i, m_j).
    \end{aligned}
\end{equation}

To formalize the objective, we define a function that minimizes the weighted sum of the three cost functions:

\begin{equation}
    \begin{aligned}
        & \min  DC + DL +  IC. \\
    \end{aligned}
\end{equation}

\subsection{Load balance}
The objective function previously presented does not take into account load balancing among the different domains. To address this issue, we propose a new metric that quantifies load balancing for the resources in the system. Given that the bandwidth within a domain is assumed to be sufficient, and RAM is relatively inexpensive compared to CPU, our focus is on the load balance of the CPU resource among different domains.

To achieve load balancing, we assume that CPU resources are provided by the number of cores, and that resources assigned to one network slice are dedicated and isolated. We use $num(m)$ to denote the number of CPUs occupied by existing slice requests in domain $m$. First, we count the total number of CPUs required by deployed network slices and incoming slices, denoted by $total$, by traversing the deployment history.

Next, we calculate the ratio, $r_m$, between the number of Virtual Network Functions (VNFs) hosted by domain $m$ and the total number of VNFs for every domain. This produces a distribution that we use to compute entropy $H$, which serves as the metric for load balancing. A higher value of $H$ indicates a better load balance for the system.

The equations used for this metric are as follows:
\begin{equation}
\begin{aligned}
    & total = \sum_{m} num(m) + \sum_{n} CPU_n \\
    & r_m = (num(m)+\sum_{n \in N} X_n^m \times CPU_n) / total \\
    & H = -(\sum_{m\in M}  r_m \times \log(r_m))
\end{aligned}
\end{equation}

In certain scenarios, it may not be desirable to distribute Virtual Network Functions (VNFs) evenly across all domains. For instance, it may be preferable for the Cloud domain to shoulder a larger proportion of the load. In such cases, instead of comparing to a uniform distribution, a predefined distribution $P$ can be established, and load balancing can be optimized based on the Kullback-Leibler (KL) divergence between the current distribution and the predefined distribution.

The predefined distribution $P$ can be expressed as a set of probabilities $p_m$ for each domain $m \in M$. The KL divergence between the current distribution and the predefined distribution can be formulated as follows:
\begin{equation}
    \label{equ:kl}
    \begin{aligned}
    & P = \{p_m, \forall m \in M\} \\
    & KL = -(\sum_{m\in M}  r_m \times \log(\frac{p_m}{r_m}))
\end{aligned}.
\end{equation} 
This approach allows for a more flexible and targeted load balancing strategy that can accommodate specific distribution preferences or requirements.
The KL divergence, a measure of the difference between two probability distributions, is employed as an indicator of load balance. The objective of load balancing is to align the distribution of CPU demandings among the domains with a predefined distribution $P$, such that the KL divergence between the current distribution $R$ and $P$ is minimized. The closer $R$ is to $P$, the lower the KL divergence will be.

To align the distribution with a discrete uniform distribution, $p(m)=\frac{1}{|M|}$ can be set. The load balancing problem can be formulated as an optimization problem with the following objective function:
\begin{equation}
    \begin{aligned}
        & \min \alpha DC + \gamma DL + \beta IC + \delta KL
    \end{aligned},
\end{equation}
where $\delta$ is a weighting coefficient that reflects the relative importance of load balancing.

The four objective functions, namely DC, DL, IC, and KL, have different scales, which may result in an imbalance during optimization. To address this issue, we utilize the min-max algorithm to standardize DC, DL, and IC.
\begin{equation}
    \begin{aligned}
        & DC_{min} = |N| \times (c_{CPU}(m_i) + c_{RAM}(m_j) \\
        & \text{where } m_i = \text{ argmin } c_{CPU}(m) \text{ for } m \in M, \\
        & m_j = \text{ argmin } c_{RAM}(m) \text{ for } m \in M.
    \end{aligned}
\end{equation}
\begin{equation}
    \begin{aligned}
        & DC_{max} = |N| \times (c_{CPU}(m_i) + c_{RAM}(m_j) \\
        & \text{where } m_i = \text{ argmax } c_{CPU}(m) \text{ for } m \in M, \\
        & m_j = \text{ argmax } c_{RAM}(m) \text{ for } m \in M.
    \end{aligned}
\end{equation}
$$$$
\begin{equation}
    \begin{aligned}
        \widehat{DC} = \frac{DC-DC_{min}}{DC_{max}-DC_{min}}
    \end{aligned}
\end{equation}
We adopt a similar normalization strategy for $DL$ and $IC$, resulting in $\widehat{DL}$ and $\widehat{IC}$. Although the theoretical value of KL divergence can be infinitely large, empirical evidence suggests that it typically ranges around 1. If the value of KL divergence is too large, then it is crucial to minimize it in order to align the distribution with the predefined distribution. Hence, the final objective function that we utilize is as follows:
\begin{equation}
    \begin{aligned}
        & \min \alpha \widehat{DC} + \beta \widehat{DL} + \gamma \widehat{IC} + \delta KL ,
    \end{aligned}
\end{equation} where $\alpha$, $\gamma$, $\beta$, and $\delta$ are hyperparameters used for generating partition plans with different characteristics. 
\subsection{Constraints}
First, the indicator variable $X_n^m$ can only take values of $1$ or $0$.
Second, each function can be assigned to only one domain.
Third, if there is an edge between node $i$ and node $j$, then the index of the domain to which node $i$ is assigned should be less than or equal to the index of the domain to which node $j$ is assigned.
This is because we assume that the VNF-FG is a Directed Acyclic Graph (DAG).
We formalize the optimization problem as follows:
\begin{equation}
    \begin{aligned}
        & \min \alpha \widehat{DC} + \gamma \widehat{DL} + \beta \widehat{IC} + \delta KL \\
        & \text{s.t.} \quad X_n^m \in \{0,1\} \quad \forall n \in N \text{ and } 1<=m<=M \\
        & \quad \sum_{n \in N} \sum_{m \in M} X_{n}^{m} = 1\\
        & \quad X_{n_i}^{m_i} \times X_{n_j}^{m_j}=0 \quad \forall (n_i, n_j) \in E \text{ and } m_i > m_j \\
        & \quad \sum_{n \in N} \sum_{m \in M} X_{n}^{m} \cdot l(m) <= l
    \end{aligned}.
\end{equation}

\section{Methodology}
\label{sec:methodology}
\subsection{Heuristic}
\subsubsection{Approximate Integer Linear Programming}
To linearize the quadratic terms in the objective function and constraints, we employ a binary integer quadratic programming (BIQP) reformulation. The fundamental approach involves introducing additional binary variables and linear constraints to replace the quadratic terms present in the original constraints and the objective function.

Another challenge that arises is the linearization of the KL divergence term, specifically the $r_m \log r_m$ component in Equation \ref{equ:kl}. In order to approximate $r_m \log r_m$, we utilize the second-order Taylor expansion of $r_m \log r_m$ at $r_m=a$:
\begin{equation}
r_m \log r_m \approx a \times \log (a) + (r_m-a) \times (\log (a)+1) +
\frac{(r_m-a)^2}{2 \times a}.
\end{equation}
For our project, we set $a=0.3$.

\subsubsection{Integer Programming: branch and bound}
Since the approximation of KL divergence sacrifices accuracy, we also employ another heuristic algorithm: branch-and-bound, to solve the problem. The fundamental idea of branch-and-bound is to divide the problem into smaller subproblems and use bounds to eliminate those subproblems that cannot possibly contain the optimal solution. The branch-and-bound algorithm can be enhanced using various techniques such as cutting planes, heuristics, and symmetry breaking. While branch-and-bound is a powerful technique for solving integer programming problems, it can be computationally expensive for large-scale problems or those with complex constraints.
\subsection{Graph Neural Network} 
\subsubsection{Model}
The linear approximation entails the optimization of $O(|V|^2)$ variables, resulting in a high time complexity for the branch and bound algorithm without guaranteed performance. As the global coordinator periodically devises the partition plan, the VNF-FG exhibits a similar structure. Consequently, we introduce a machine learning algorithm utilizing a graph neural network (GNN) for partitioning the VNF-FG, which we refer to as GNNP.

A fundamental framework of GNN is message passing. Each node in the graph is initialized with its feature vector, constituting the first layer. Subsequently, each node gathers information from its neighboring nodes, generating a new vector. These newly generated vectors form the second layer. By following this procedure, $k$ layers are produced, where $k$ is a hyperparameter that determines the GNN model's complexity. Numerous aggregation methods are available; however, we opt for the Graph Convolutional Network (GCN) \cite{kipf2016semi}. Generally, it is assumed that the vectors from the final layer encompass all topological information. Therefore, we transform the vectors from the last layer into the designated domain. The comprehensive algorithm is presented in Algorithm \ref{alg:gnn_partitioner}.

\begin{algorithm}[tb]
\caption{GNN Partitioner}
\label{alg:gnn_partitioner}
\begin{algorithmic}[1]
\REQUIRE ~~\\
    VNF-FG $G=(N, E)$, List of Domains $M$, the number of layers $K$
\ENSURE 
    Assignment for each node $n \in N$
\STATE Initialize the feature vector for each node i $x_i^{(0)}$
\FOR{k in {1,..,K}}
    \STATE 
        $$
            x_i^{(k)} = \sum_{j \in \mathcal{N}(i) \cup \{i\}} \frac{1}{\sqrt{\deg(i)} \cdot \sqrt{\deg (j)}} \cdot (W^\top \cdot x_j^{(k-1)}+b)
        $$
\ENDFOR
\STATE Embedding vector for node i: 
    $$x_i = \text{SoftMax} (W \cdot x_i^{K}+b)$$
\STATE $X_i^m$ = the $m^{th}$ element of $x_i$ 
\RETURN $X_i^m$
\end{algorithmic}
\end{algorithm}

\subsubsection{Optimization} 
\begin{figure}[tb]
\centerline{\includegraphics[width=0.5\textwidth]{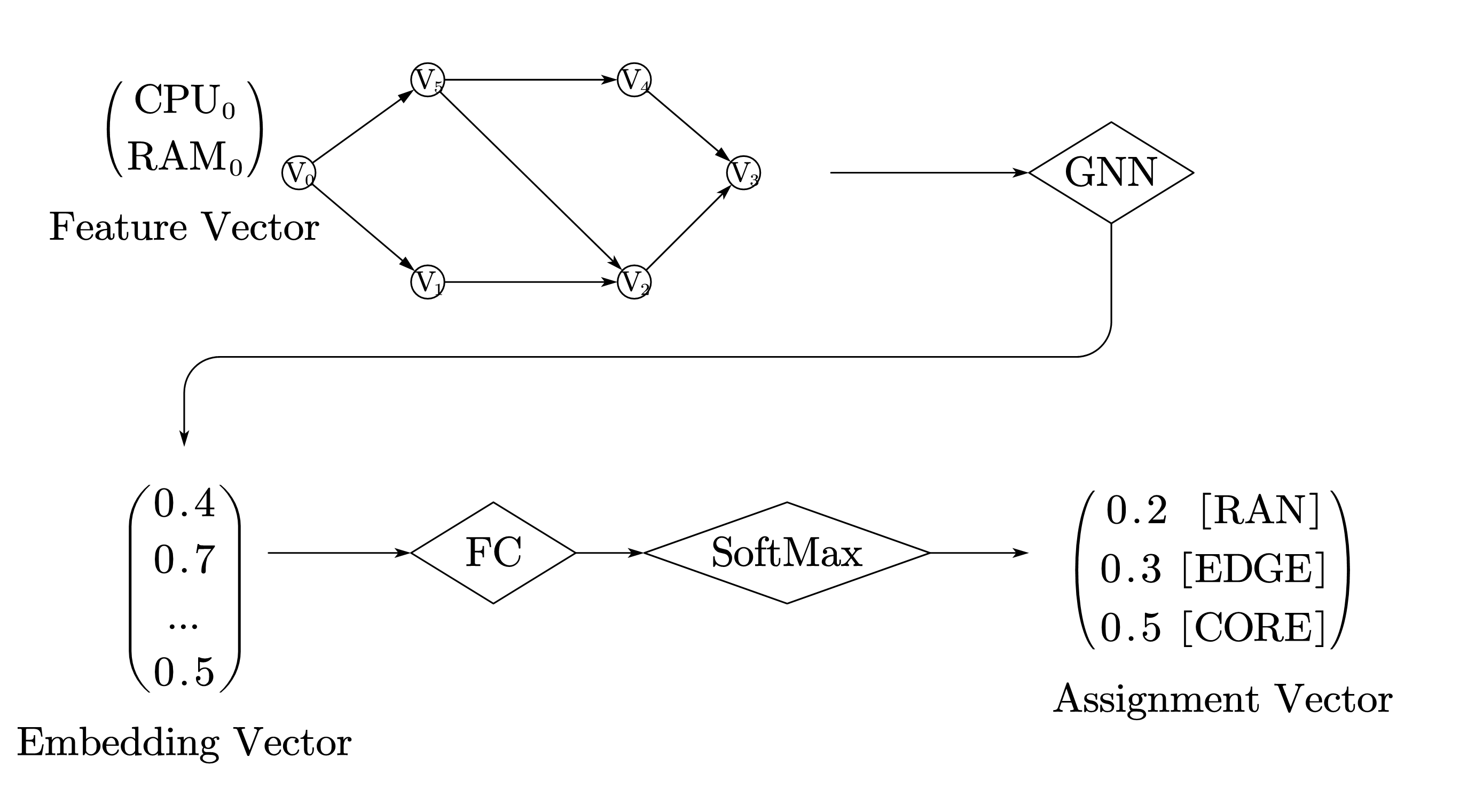}}
\caption{GNN Partitioner}
\label{fig:GNN_procedure}
\end{figure}

As the range of the $x_i$ cannot be directly controlled, constraints are incorporated into the objective function by employing relatively large coefficients. The first constraint is relaxed from a binary variable to a continuous variable ranging from 0 to 1. By virtue of the SoftMax function's definition, the second constraint is inherently satisfied. Consequently, the objective function can be formulated as follows:
\begin{equation}
    \begin{aligned}
        \min \quad & \alpha \widehat{DC} + \gamma \widehat{DL} + \beta \widehat{IC} +  \delta KL + \\ 
        & \mu (X_{n_i}^{m_i} \times X_{n_j}^{m_j} \quad \forall (n_i, n_j) \in E \text{ and } m_i > m_j),
    \end{aligned}
\end{equation}
where $\mu$ is a sufficiently large value to ensure that the solution satisfies the third constraint. 
Given that this is not a conventional supervised learning task, backpropagation need only be performed based on the direct loss value. The complete procedure is depicted in Fig. \ref{fig:GNN_procedure}.
\section{Experiments}
\label{sec:experiments}
\subsection{Environment Setup}
\begin{figure}[tb]
\centerline{\includegraphics[width=0.3\textwidth]{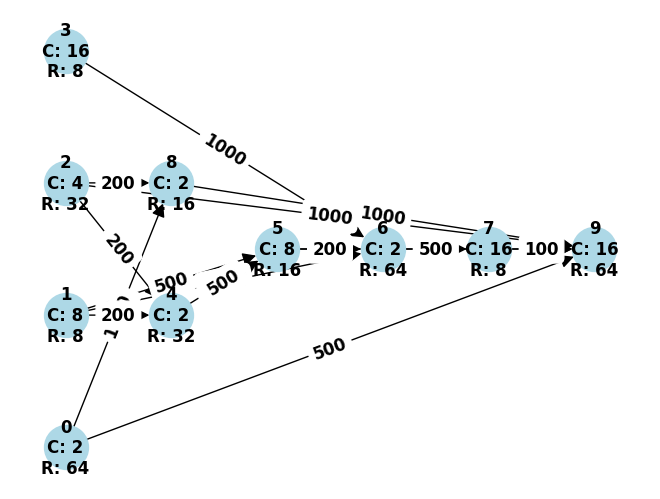}}
\caption{Example of VNF-FG}
\label{fig:VNFFG}
\end{figure}

To represent 200 slices' VNF-FG, we generated 200 random DAGs. Each graph had the freedom to select its number of nodes from the set $[10, 15, 20]$ and its number of edges from the set $[15, 30, 60]$. Each node could choose its CPU level from the set $[2, 4, 8, 16]$ and its RAM level from the set $[8, 16, 32, 64]$. Each edge had the flexibility to select its bandwidth level from the set $[100, 200, 500, 1000]$. An example of the DAG is depicted in Figure \ref{fig:VNFFG}.
To train the GNNP model, we generated an additional 800 DAGs.

We created four domains: RAN domain, edge domain, core domain, and cloud domain, each with decreasing resource costs per unit. The CPU costs per unit for these domains were $100$, $50$, $20$, and $5$ respectively, while the RAM costs per unit were $10$, $5$, $2$, and $1$, respectively. The bandwidth cost per unit were $1$, $0.5$, $0.2$, and $0.1$ for RAN, edge, core, and cloud domains respectively. Additionally, we assigned an inter-domain cost per unit as 10, 5, and 2 for domain pairs (RAN, edge), (edge, core), and (core, cloud), respectively. The target CPU distribution for these domains was 0.1, 0.2, 0.3, and 0.4 for RAN, edge, core, and cloud domains respectively.

\subsection{Metrics}
\subsubsection{Total Resource Cost}
Due to the significant variations in cost between different VNF-FGs, which may differ in the number of nodes and edges, we chose to compare algorithms based on the summation of costs for all 200 VNF-FGs, rather than their average cost.
\subsubsection{Divergency from Target Distribution}
The total resource cost calculation does not incorporate any deviation from the target distribution. Our objective is to evaluate the degree of alignment between the distribution generated by algorithms and the target distribution.
\subsection{Evaluation}
\subsubsection{GNNP Model Selection}
We partitioned the supplementary 800 VNF-FGs into two distinct sets: 700 for training and 100 for validation purposes. The subsequent hyperparameter tuning procedure was conducted under a fixed set of parameters, specifically $\alpha$, $\beta$, $\gamma$, and $\delta$.
\begin{itemize}
    \item Latent Variable Size: Increasing the size of the latent variable results in a higher number of parameters within the model. Given that our initial feature dimensionality is only 2, it is prudent to avoid using an excessively large latent variable size. In our study, we evaluated the performance of the model with latent variable sizes of 5, 10, 25, and 50. The number of layers is fixed to $3$. The results are shown in Fig. \ref{fig:latent_variable_size}\footnote{\label{note1} All of the boxplots in this paper are drawn based on 50 experiments. The circles denote the outlier. The upper bound of the vertical line, the upper bound of the box, the middle horizontal line within the box, the lower bound of the box, and the lower bound of the vertical line represent the minimum, the first quartile, the median, the third quartile and maximum, respectively.}. 
    \begin{figure}[tb]
    \centerline{\includegraphics[width=0.5\textwidth]{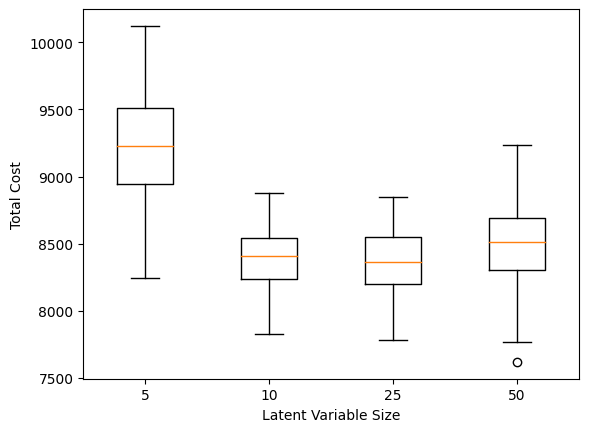}}
    \caption{Latent Variable Size Tuning}
    \label{fig:latent_variable_size}
    \end{figure}
    Based on the experimental outcomes, a latent variable size of 10 appears to strike an optimal balance between model stability, performance, and computational complexity. 
    \item The number of layers, denoted as $k$, plays a crucial role in determining the performance and complexity of the model. As $k$ increases, each node attains a more comprehensive global perspective, which in turn augments the model's complexity. We experiment with four distinct values for $k$: 1, 3, 5, and 7. The outcomes of these experiments are illustrated in Figure \ref{fig:num_layers}.
    \begin{figure}[tb]
    \centerline{\includegraphics[width=0.5\textwidth]{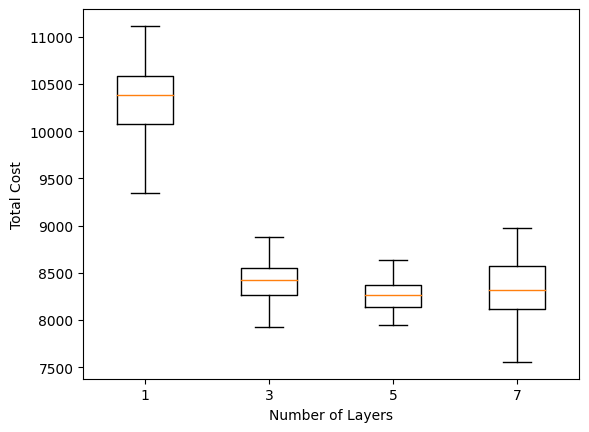}}
    \caption{Number of Layers Tuning}
    \label{fig:num_layers}
    \end{figure}
\end{itemize}
A prevalent observation indicates that, as the complexity of the model increases, there is a corresponding rise in average performance. Nonetheless, this increased complexity may also contribute to overfitting and an escalation in variance, resulting in diminished stability.

\subsubsection{Algorithm Comparasion}
We conduct a comparative analysis of three algorithms: Graph Neural Network for Partitioning (GNNP), Approximate Integer Linear Programming (AILP), and Branch and Bound (BnB). The results are presented in Table \ref{table:algo_compara}. AILP demonstrates the most favorable average total cost, but it exhibits the least congruence with the target distribution. This discrepancy could be attributed to the approximation of the Kullback-Leibler (KL) divergence within the objective function. In contrast, BnB necessitates the longest inference time, and its performance is not guaranteed. GNNP produces outputs that align closely with the target distribution and boasts the shortest inference time. However, GNNP demands a comparatively extended training duration.

\begin{table}
\centering
\caption{Algorithm Comparasion}
\label{table:algo_compara}
\begin{tabular}{|l|l|l|l|} 
\hline
                   & GNNP          & AILP          & BnB   \\ 
\hline
Total Cost         & 8539          & \textbf{8316} & 9539  \\ 
\hline
KL-divergence      & \textbf{0.05} & 0.18          & 0.08  \\ 
\hline
Inference Time (s) & \textbf{0.2}  & 2.4           & 5.3   \\
\hline
\end{tabular}
\end{table}
\subsection{Ablation Study}
\subsubsection{Effects of Standardization Cost Items in Objective Function}
\begin{figure}[tb]
\centerline{\includegraphics[width=0.4\textwidth]{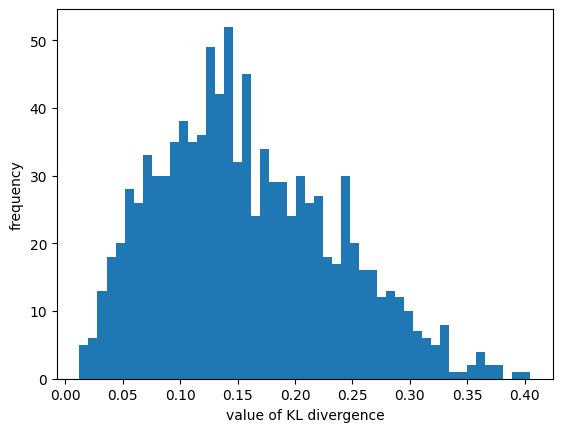}}
\caption{KL divergency value distribution}
\label{fig:kl_div_dist}
\end{figure}

Initially, we must verify the scale of the Kullback-Leibler (KL) divergence. We sample $10^4$ random distributions and calculate the KL divergence between them and the target distribution. The distribution of KL divergence values is depicted in Figure \ref{fig:kl_div_dist}.
In order to align the KL divergence item with other uniformed components in the objective function, we scale up the KL divergence item by a factor of two. Utilizing the GNNP algorithm, the KL divergence value is $0.249$ before the adjustment and $0.186$ after the adjustment.

If we do not standardize other terms in the objective function, the impact of the KL divergence item becomes negligible. This is due to the larger scale of other terms compared to the value of KL divergence. Without standardizing the other items, the average cost of partition plans generated by the GNNP algorithm is $8492$, and the KL divergence is $0.58$.

\subsubsection{Effect of $\alpha$, $\beta$, $\gamma$, $\delta$}
By constraining the value of each term between 0 and 1, we can adjust the coefficients of individual items to generate different partition plans with varying tendencies. We provide two examples, both generated using the GNNP algorithm.

In the first example, we double the coefficient for the CPU cost. Before doubling, the average CPU cost is $3217$, which drops to $2793$ after doubling. Simultaneously, the ratio of CPU in the cloud domain increases from $0.51$ to $0.56$ after the adjustment. 

In the second example, we double the coefficient for the inter-domain bandwidth cost. The resulting distribution becomes $[0.17, 0.31, 0.35, 0.17]$. As observed, most nodes are situated in the first three domains to lower the inter-domain cost, and the KL divergence is $0.134$.
\section{Conclusion}
In this paper, we explore the problem of multi-domain end-to-end network slice partitioning. We define resource costs and incorporate load balancing among different domains into the objective function. We propose two heuristic algorithms and one machine learning-based algorithm, comparing them using various metrics. Additionally, we tune the weights for different terms and demonstrate their effects. By adjusting the assigned weights, we can generate partition plans with distinct characteristics.
The latency constraints were not considered in our algorithm. To incorporate them, the domains' ability to offer latency for each type of VNF can be assumed, and a constraint can be imposed on the total processing delay. This constraint can be treated as another cost element in the objective function. Our algorithm can also handle hard assignments where certain VNFs can only be assigned to specific domains. In such cases, the indicator variable for the VNF can be set to 1 for the assigned domain and $0$ for all other domains without compromising the original framework.
In future work, we aim to integrate the intelligent partitioning algorithm with our proposed multi-domain network slice provisioning architecture, to achieve more efficient resource utilization.
\label{sec:conclusion}
\section*{Acknowledgment}
This work was supported in part by the National Science Foundation under Grant No. CNS-2008856.
\bibliographystyle{ieeetr}
\bibliography{reference}
\end{document}